\documentstyle[preprint,aps,psfig]{revtex}
\draft
\tighten

\newcommand{\vecr}{{\bf r}}
\newcommand{\vecR}{{\bf R}}
\newcommand{\half}{\frac{1}{2}}

\begin{document}

\title{Multi-step Effects in Subcoulomb Breakup}
\date{\today}

\author{F.M. Nunes\footnote{Email address: filomena@wotan.ist.utl.pt}}
\address{$^{1)}$Universidade Fernando Pessoa, Pra\c{c}a 9 de Abril, 4200 Porto, Portugal \\
$^{2)}$CENTRA, Instituto Superior T\'ecnico, 1096 Lisboa-Codex, Portugal.}  

\author{I. J. Thompson}
\address{Department of Physics, University of Surrey, Guildford GU2 5XH, UK}  
\maketitle

\begin{abstract}
Following earlier one-step calculations, we explore the
contributions of multi-step effects for the breakup of low energy $^8$B
on $^{58}$Ni  and $^{208}$Pb within a coupled discretised continuum channels (CDCC) formalism.
The Coulomb multi-step differential cross section is significantly
reduced for all angles, the largest effect being
the destructive interference of nuclear couplings.
The nuclear peak, at around $80^\circ$ in the one-step
calculations for $^{58}$Ni, virtually disappears.
\end{abstract}

\section{Introduction}

Coulomb dissociation has been proposed on many occasions
\cite{shotter,baur} as a means of determining the interaction between
fragments at low relative energies. This method is applicable 
even when the fragments are
themselves radioactive and not easily produced as targets for direct
scattering experiments.  It was believed that at sufficiently forward
angles, and/or at sufficiently low energies, the impact parameters for
the breakup trajectory would be large enough for Coulomb mechanisms to
dominate, and for first-order theories of Coulomb breakup to be
adequate. The Coulomb dissociation method deduces the radiative
capture cross section by measuring the reverse reaction, the dissociation of a projectile 
(the fused system) by the Coulomb field of a target.

The Coulomb dissociation method has been used to examine the breakup of
$^8$B at both high \cite{moto1,moto2} and low \cite{nd} energies.
Analyses have started with semiclassical theory \cite{alder}, and have
progressed to include the $E1/E2/M1$ contributions with correct
experimental efficiencies \cite{shyam96}, three-body kinematics in the
final state \cite{shyam97}, and most recently one-step nuclear and
Coulomb contributions \cite{nunes1,puri,shyam98}.  These last results,
and those of ref.  \cite{vitturi}, showed that the nuclear and Coulomb
form factors extend to considerably larger distances than the sum of
the radii of the participating nuclei, because of the extended tail of
the wave function of the last proton in  $^8$B, and that there is
strong Coulomb-nuclear destructive interference at intermediate radii.
We are prompted, by the size of these effects, to examine the
importance of {\em multi-step} contributions for both Coulomb and
nuclear processes, taking into account the final state interactions
which were previously omitted.  These final-state interactions will
couple together the different continuum states, and also describe the
depletion of the elastic channel due to breakup. The depletion effect
has been considered in some calculations \cite{vitturi}, where only
couplings between the bound state and continuum states were included.
The contribution of `higher-order breakup' has yet to be properly clarified,
especially the role of continuum-to-continuum couplings.

Earlier treatments of multi-step effects for breakup have used either
adiabatic \cite{thom83} or semiclassical approximations
\cite{bertu1,roman}, solved the scattering problem with time-dependent
methods \cite{esb}, tried CDCC solutions \cite{saku,cdcc}, or used
Bremstrahlung integrals \cite{jat1,jat2}. Estimates of second-order
Coulomb and nuclear effects have also been calculated \cite{typel}.
  
The earliest adiabatic approximation used the three-body wave functions
of Amakawa et al. \cite{amakawa} within a prior-form breakup matrix
element \cite{thom83} for both nuclear and Coulomb mechanisms in the
$^7$Li breakup into $\alpha +t$, when incident on $^{208}$Pb at 70 MeV.
However, the best fit to the experimental cross sections was found when
both Coulomb distortion and Coulomb breakup were omitted from the
calculation.
The breakup  of $^6$Li into $\alpha +d$, when incident on $^{208}$Pb at
156 MeV, was later analysed \cite{saku} within the CDCC framework
\cite{cdcc}, and strong nuclear and Coulomb interference effects were found,
even at forward angles. We follow a similar CDCC approach,
but extend it to include, for the first time, dipole as well as
quadrupole Coulomb mechanisms.

Recent investigations of the breakup of halo nuclei have prompted a
revival of semiclassical treatments of breakup
\cite{bertu1,roman,bertu2} where the continuum is discretised into an
orthogonal set of basis functions.  First-order and higher-order
couplings can then be included when integrating along a semiclassical
trajectory (Rutherford orbit at low energies, or straight lines at high
energies). Simplified ground-state wave functions are often used, and
collective rather than semi-microscopic form factors calculated. We
will see below that both of these approximations have to be reviewed
when we consider the breakup of $^8$B incident on $^{58}$Ni at the
sub-Coulomb energy of 26 MeV.

Another  method, successfully used \cite{esb,kido96} to treat
higher-order processes, is to follow the breakup reaction as a
time-dependent process. Esbensen et al. \cite{esb} follow the time evolution
of a $^7$Be+p bound state by means of a TDHF propagator along a
straight-line trajectory, and can calculate both Coulomb and nuclear
contributions to breakup in a unitary manner. The results \cite{esb} show
that the Coulomb higher-order dynamical processes cause a destructive 
E1/E2 interference, and a reduction of the dissociation probability.

Most recently, the adiabatic  three-body wave functions have been used
again \cite{jat1,jat2} for Coulomb breakup, since in this case analytic
solutions have been discovered both for the three-body wave functions
\cite{rcj97} and for the  post-form $T$-matrix integral of the breakup matrix
element \cite{jat1} in terms of a Bremstrahlung integral. Unfortunately,
the method is not immediately applicable for
$^8$B breakup, since these analytic solutions only hold for
neutral valence particles, and  the adiabatic approximations are
for high energy rather than for sub-Coulomb reactions.

We have been progressively improving our understanding of
low energy breakup reactions \cite{nunes1,puri,shyam98}, focusing in particular
on the breakup of $^8$B on $^{58}$Ni measured by the Notre Dame
group \cite{nd} at 26 MeV. The work presented here is an exploratory continuation of
our previous investigations, and now, for the first time, a full multi-step quantum
mechanical description is attempted, including all continuum couplings.

\section{Theory}

\subsection{Coupled Discretised Continuum Channels (CDCC)}

When a projectile is described as a single particle outside a core, its
state can be disturbed by the interaction with the
target nucleus, as the tidal forces of the target act differentially on
the particle and the core. 
If one separates the projectile-target interaction into
$V_{ct} (\vecR_c)$, the interaction of the target nucleus with the core, 
and $V_{pt}(\vecr')$, the interaction of the target nucleus with the particle, 
then there is a mechanism for  coupling
ground and inelastic (continuum) states together.
Nuclear and
Coulomb components of $V_{ct}$ and $V_{pt}$ should be included on the same footing.

In order to describe the breakup of a projectile such as $^8$B, we
could consider the inelastic excitations in the $p+^7$Be system from the
ground state $\phi_{gs}(r)$ to excited states in the continuum $u_{\ell
sj,k}(r)$, for some momentum $k$ and partial wave $\ell$.  The use of
such single energy eigenstates, however, would result in calculations of
the inelastic form factors which will not converge, as the continuum
wave functions do not decay to zero as $ r \rightarrow \infty$
sufficiently fast to be square integrable.  One way \cite{CDCC,CDCC2}
of dealing with this divergence is to take continuum states, not at a
single energy, but averaged over a narrow range of energies, such that these
`bin' states {\em are} square integrable. We label these bin states
by their wave-number limits $[k_1,k_2]$ and their angular momentum quantum
numbers $(\ell s)j$. We use them in the coupled discretised continuum channels
(CDCC) method \cite{CDCC,CDCC2}.

Let $\vecR$ be the coordinate from the target to the projectile and
$\vecr$ the internal coordinate of the projectile.
The position coordinates of the projectile fragments with respect to the target are
$$  \vecr' = \vecR + \frac{A_p-1}{A_p} \; \vecr \; $$ and
$$  \vecR_c = \vecR - \frac{1}{A_p} \; \vecr \; ,$$
where the projectile has mass $A_p$.

The wave function for the three-body system of proton + $^7$Be core
+ target is expanded as
\begin{eqnarray} \label{Psieq}
 \Psi _J^M
   \left ( \vecR , \vecr \right ) =
 \sum _ {{L \ell j,[k_1,k_2] } , {M_L \mu}}
 \phi ^\mu_ {\ell sj,[k_1,k_2]} ( \vecr)
  \langle LM_L  j \mu | JM \rangle
      i ^ L  Y_L^{M_L} (\hat{\vecR} ) \frac{1}{R}
        f_{L,\ell j[k_1,k_2],J} (R) \;,
\end{eqnarray}
with
\begin{eqnarray} \label{phieq}
 \phi ^\mu_ {\ell sj,[k_1,k_2]} ( \vecr) = 
 \frac{1}{r} \sum_{m\sigma} \langle \ell m s \sigma | j \mu \rangle
  ~Y_\ell^m (\hat{\vecr}) ~\chi_s^\sigma ~u_{\ell sj,[k_1,k_2]} (r) \;,
\end{eqnarray}
where $\chi_s^\sigma$ is the proton's intrinsic state.
The set $\{L,\ell,s,j,[k_1,k_2]\}$ will be abbreviated as $\alpha$.

The radial wave functions $f _ {L,\ell j[k_1,k_2],J} (R)$
satisfy the set of coupled equations
\begin{eqnarray} \label{CRC}
 \left [ 
- \frac{\hbar^2 }{2 \mu} ~ \left ( \frac{d^2 }{ dR^2} - \frac{L(L+1)} {R^2} \right )
+ \epsilon([k_1,k_2]) - E \right ]
 f _ {\alpha J} (R)  + \sum _ {\alpha '} 
 i ^ {L ' - L} ~ V^J _{\alpha:\alpha'}(R)  f_{\alpha' J} (R) = 0,
\end{eqnarray}
where $\epsilon([k_1,k_2])$ is the average energy of continuum bin $[k_1,k_2]$,
(or  $\epsilon<0$ for the ground state). $V^J _{\alpha:\alpha'}(R)$
describes the coupling between the different relative motion states:
\begin{eqnarray} \label {CP}
V^J_{\alpha;\alpha^\prime} & = & < \phi_\alpha(\vecr)|
	V _ {ct} (\vecR _ c) + V _ {pt} (\vecr ')|\phi_{\alpha^\prime}(\vecr)>,
\end{eqnarray}
where $V_{ct} (\vecR_c) $ and $V_{pt}(\vecr')$
are again the total (nuclear and Coulomb) interactions between
$c-t$ and $p-t$ systems respectively.
In Eq.(\ref{CP}) radial integrations are done over $r$  
from zero to $R_{bin}$, a parameter to be chosen.

The coupled equations of Eq.(\ref{CRC}) may be solved exactly \cite{fresco}
if they are not too numerous. Otherwise, iterative expansions are used 
starting with $f^{(-1)}(R)=0$, and continuing as
\begin{eqnarray}  \nonumber
 &~& \left [ 
   -\frac{\hbar^2 }{2 \mu} ~ \left ( \frac{d^2 }{ dR^2} - \frac{L(L+1)} {R^2} \right )
   + V^J _{\alpha:\alpha}(R)  + \epsilon([k_1,k_2]) - E \right ]
 f^{(n)} _ {\alpha J}(R)  \\
  & ~ & \hspace{2cm} = \sum _ {\alpha'\ne\alpha} 
 i ^ {L ' - L} ~ V^J _{\alpha:\alpha'}(R)  f^{(n-1)}_{\alpha'J} (R)\;,
\label{DWBA}
\end{eqnarray}
for $n$=0, 1, ... The function $f^{(0)}(R)$ is thus the elastic
channel, and the asymptotic $S$-matrix $S^{(n)}$ of the wave functions
$f^{(n)} (R)$ gives the cross section for $n^{th}$-order DWBA. 
The $n=1$ first-order DWBA solutions are presented in the previous
paper \cite{shyam98}. The multi-step DWBA results for large $n$ 
will converge to the coupled-channels solution
if the off-diagonal couplings are small. If they are large, the DWBA
series diverges. Then the infinite series may be summed by the method
of Pad\'e approximants described below.

The bin wave functions are defined as
\begin{equation} \label{binwf}
{  u_{\ell sj,[k_1,k_2]} (r) = \sqrt {\frac{2 }{\pi N}} ~~
            \int _ {k _ 1} ^ {k _ 2} w(k) e^{-i\delta_k} u _{\ell sj,k} (r) dk}\;,
\end{equation}
with $\delta _ k$ the scattering phase shift for $ u _ {\ell sj,k} (r) $,
the single-energy scattering wave function in the chosen potential $V^\ell_{pc}(r)$
which may be $\ell$-dependent.
The normalisation constant is $N = \int _ {k _ 1} ^ {k _ 2} |w(k)| ^ 2 dk $
for the assumed weight function $w(k)$, here taken to be unity.
These bin states are normalised $ \langle u | u\rangle = 1 $
once a sufficiently large maximum radius $R_{bin}$ for $r$ is taken.
They are orthogonal to any bound states, and are orthogonal to other bin
states if their energy ranges do not overlap.
The phase factor  $e^{-i\delta_k} $ ensures that they are all real valued
for real potentials $V^\ell_{pc}(r)$.

The rms radius of a bin wave function increases as the bin width
$k_2-k_1 $ decreases, approximately as $1/(k_2-k_1)$, so large radial
ranges are needed to include narrow bin states.  If the maximum radius
$R_{bin}$ is not sufficiently large, then the bin wave functions 
$u_{[k_1,k_2]}$ will
not accurately be normalised to unity by the factors given in equation
(\ref{binwf}). It is important however, to realise that the missing
normalisation comes at large distances; the bin wave functions must not
be artificially renormalised to unity, otherwise, for example, the
correct Coulomb B(E$K$) distributions will not be obtained.

The couplings $V^J_{\alpha:\alpha'}(R)$ in Eq.(\ref{CRC})
arise, as discussed above,
from the interaction potentials of the projectile fragments with the target.
Assuming that the potentials $V _ {ct}$ and $V_{pt}$ are central,
the Legendre multipole potentials can be formed as
\begin{equation}
  {\cal V} _ K (R,r) = \half \int _ {-1} ^ {+1} 
    \left [ V _ {ct} (\vecR _ c) + V _ {pt} (\vecr ') \right ]
       P _ K (x) dx \;,
\end{equation}
where 
$K$ is the multipole and $x = \hat{\vecr}\cdot \hat{\vecR} $
is the cosine of the angle between $ \vecr $ and $ \vecR$.
Since $s$ is the (fixed) spin of the proton, a spectator,
the coupling form factor between states $u _ {\ell'[k_1,k_2]'} (r) $
and $u _{\ell[k_1,k_2]} (r) $ is then
\begin{eqnarray} \nonumber
 V^J _{\alpha : \alpha '} (R)
     & =&  \sum _ K  ~
  (-1)^{j+j'-J-s} \hat{j} \hat{j'} \hat{\ell} \hat{\ell'} \hat{L} \hat{L'} 
       (2K+1) W(jj'\ell\ell';~K s)~W(jj'LL';~K J) \\
&\times& \left ( \begin{array}{ccc} K & \ell & \ell'   \\
                           0 & 0    & 0 \\ \end{array} \right )
         \left ( \begin{array}{ccc} K & L & L'   \\
                           0 & 0 & 0 \\ \end{array} \right )
     { \int _ 0 ^ {R_{bin}} ~u _ {\ell sj[k_1,k_2]} (r)
                ~{\cal V} _ K (R,r) ~u _ {\ell'sj'[k_1,k_2]'} (r) ~dr} \;.
  \label{binint}
\end{eqnarray}

 From the $S$-matrices $S^{(n)}$ we calculate the double differential
cross sections $d^2\sigma/d\Omega d\epsilon$, where $\Omega$ is the
scattering angle of the centre of mass of the $^8$B$^*$ fragments $p$
and $c$, and $\epsilon$ is the excitation energy within $^8$B$^*$.
Usually we will plot the integrated angular distribution
$d\sigma/d\Omega$ obtained after summation over all the bin energies
$\epsilon([k_1,k_2])$. It will be possible \cite{CDCC2,CDCC3} to obtain
from the CDCC results the full multiple differential cross sections
such as $d^3\sigma/d\Omega_p d\Omega_c dE_p$, and then, for example,
any post-acceleration effects may be determined from the variation of
the cross section with respect to the laboratory proton energy $E_p$.
An approximate treatment is available \cite{shyam97,shyam98} to calculate these
fragment distributions using an isotropic assumption for the $pc$ relativ
motion, but this does not yet include the interference
between the different final $\ell$-states that is necessary
\cite{roman} to give non-zero post-acceleration effects.

\subsection{Pad\'e acceleration}
\label{pade}

A given sequence $S^{(0)} , S^{(1)} , \cdots $ of S-matrix elements
that result from iterating the coupled equations
can be regarded as the successive partial sums of the polynomial
\begin{eqnarray}
  f(\lambda) = S^{(0)} + (S^{(1)} - S^{(0)}) \lambda
                   + (S^{(2)} - S^{(1)}) \lambda ^ 2 + \cdots
\end{eqnarray}
evaluated at $\lambda$=1.
This polynomial will clearly converge for $\lambda$ sufficiently small,
but will necessarily diverge if the analytic continuation of the
$f (\lambda) $ function has any pole or singularities inside the circle
$ | \lambda |<1 $ in the complex $\lambda$-plane.
The problem that Pad\'e approximants solve is that of finding a computable
approximation to the analytic continuation of the $f (\lambda) $ function
to $\lambda$=1.
This is accomplished by finding a rational approximation
\begin{eqnarray}
 P_{[N,M]} (\lambda) = {{p_0 + p_1 \lambda + p_2 \lambda^2 + \cdots + p_N \lambda^N }
                  \over
                   {1       + q_1 \lambda + q_2 \lambda^2 + \cdots + q_M \lambda^M }}
\end{eqnarray}
which agrees with the $f(\lambda)$ function in the region where the latter
does converge, as tested by matching the coefficients in the polynomial
expansion of $ P_{[N,M]} (\lambda) $ up to and including the coefficient of
$ \lambda^{n} $ for $n=N+M$.

There are many different ways \cite{graves} of evaluating the coefficients
$p_i , q_j$, but for the present problem we can use
Wynn's $\epsilon$-algorithm \cite{wynn,genz}, which is a method of evaluating
the upper right half of the Pad\'e table at $\lambda$=1 directly
in terms of the original sequence  $S^{(0)} , S^{(1)} , \cdots $.
Experience has shown that for typical sequences the most accurate Pad\'e
approximants are those near the diagonal of the Pad\'e table.
We use $\overline{S^{(n)}} \equiv P_{[N,M]} (1)$ for $N=[(n+1)/2]$ and $M=[n/2]$
in calculating the Pad\'e-resummed cross sections.

When accelerating a {\em vector} of S-matrix elements
$ {\bf S}^{(n)}$, with a component for each coupled channel $\alpha$,
then it is important to accelerate the vector as a whole.
Wynn \cite{wynn2} pointed out that this can be done using the Samuelson inverse
\begin{eqnarray}
  {\bf x} ^ {-1} = ({\bf x} \cdot {\bf x} ^ * ) ^ {-1} {\bf x} ^  * 
\end{eqnarray}
where $ {\bf x} ^  *  $ is the complex conjugate of ${\bf x}$.

\section{Results}

The multi-step DWBA calculations presented here were calculated using
{\sc Fresco} \cite{fresco}. We use a continuum breakup subspace sufficient to
reproduce what we believe to be the principal channels.
For the distorted waves of the projectile-target wave function, radii up
to  $R_{coup}=300$ fm and partial waves up to $L_{max}=600$ were
included to ensure full convergence of the individual angular
distributions. These limits give, by semiclassical considerations,
cutoffs for Coulomb excitations below 2.0$^\circ$ from our $L_{max}$,
and below 1.7$^\circ$ from our $R_{coup}$ value.  We
have examined the convergence with respect
to $R_{bin}$. For the case we are interested in, the differential
cross section remains unaltered as long as $R_{bin} \ge 50$ fm. 
Thus, each energy bin (Eq.\ref{binwf} and Eq.\ref{binint}) 
is calculated using $R_{bin}=50$ fm.

It is essential for physical completeness that our calculations include 
monopole, dipole and quadrupole contributions for both nuclear and Coulomb
mechanisms. However, they do not include $M1$ transitions. At 
the extreme nonrelativistic velocities of interest here, these are predicted 
(see for instance \cite{nd}) to be insignificant.

In Fig.(\ref{fig:bin})  we show the energy distribution
of the cross section obtained within a 1-step calculation, using the
$^8$B model from Esbensen \cite{esb}. We keep the same p+$^7$Be
potential (that defined for the ground state) for all partial waves of relative motion.
For scattering from $^{58}$Ni we use the same optical
potentials \cite{moroz,becc} as in \cite{nunes1}. The cross section is
plotted as a histogram to illustrate the continuum discretisation
that we have used to define the energy bins included in all
calculations.  These results show that transitions from the $^7$Be-p
$p_{3/2}$ ground state to s,p,d, and f-wave continuum states up to $3 $ MeV
should be taken into account, even though one can expect f-waves to
offer only a small correction to the overall result. A finer discretisation
of p, d and f waves would be desirable, but the present (70 bin)
calculation is already at the limit of our computational capacity.

We first calculate multi-step effects by iterating the
coupled equations beyond the first-order DWBA. We find that even
the second-order DWBA diverges rapidly, especially for low partial
waves (small impact parameters), and does not give sensible results beyond
30$^\circ$ (dashed line with circles in Fig. \ref{fig:steps}). In
order, therefore, to present some indications on what may be deduced
from the successive Born terms, we will present the results when
resumming the expansion using the method of Pad\'e acceleration (see section
\ref{pade}) for successive numbers of steps $n$.
We will use for cross sections the Pad\'e approximants the $\overline{S^{(n)}}$ 
rather than the original $S^{(n)}$ matrices.
Given the nature of our expansion,  it is not possible
to directly compare 2nd and 3rd order effects with those obtained using
the pure conventional DWBA expansion.  

Fig.(\ref{fig:steps}) shows the
1-step, 2-step, 3-step, 6-step and 20-step breakup results using Pad\'e
acceleration. The rate of convergence of this resummed expansion is
encouraging, contrary to that of the original Born series which diverges
strongly immediately at 2nd order.  In addition, we show in
figs.(\ref{fig:stepc}) and (\ref{fig:stepn}) the different rate of
convergence for the $^8$B breakup into s and p continuum states
considering the Coulomb  and nuclear interactions separately.  In all
the cases we have studied, the Pad\'e convergence is non monotonic.  If
one includes 1-step and 2-step processes only, the differential cross
section is underestimated.  Introducing 3-step corrections
overestimates the cross section. From our results we conclude that for
the breakup of low energy $^8$B on $^{58}$Ni, contributions up to at
least 9th order in the Pad\'e expansion should be included.

Still in Fig.({\ref{fig:steps}) we present the results obtained for the full coupled
channel
calculation taking into account s-wave and p-wave bins. We find that processes
beyond 20-step do not contribute to the cross section. For this
reduced bin subspace (42 bins) it is possible to perform the 
full coupled channel
calculation (light solid curve in Fig.\ref{fig:steps}). It is reassuring
to find that our results using the multi-step expansion with Pad\'e acceleration 
converge to the correct full CDCC results. In the larger bin
subspaces it is extremely hard to perform the full CDCC calculation
and thus we will rely on the multi-step expansion with Pad\'e acceleration.

Including multi-step effects, the Coulomb differential cross section
is hardly modified up to $\theta \simeq 10^\circ$. 
The peak at $\theta \simeq 20^\circ$ is shifted to slightly smaller angles 
with higher order processes and its magnitude is reduced by  
$\simeq 10\%$  (see Fig.\ref{fig:stepc}). 

The most striking result of our work is clearly the destructive interference 
caused by the nuclear multi-step processes. The nuclear peak is shifted to lower 
angles (from $\theta \simeq 80^\circ$ for 1-step calculations to 
$\theta \simeq 40^\circ$ for the CDCC calculations) and suffers a reduction
to $\frac{1}{6}$ of its peak value. We do not expect measurements of the 
breakup differential cross section at larger angles to provide
a good handle for the optical potentials as one could deduce
from the 1-step results presented in our earlier work \cite{nunes1}. 
The previously observed strong nuclear peak is practically 
washed away by multi-step effects.

We point out that, as in the 1-step calculations of \cite{nunes1},
the total differential cross section does  not correspond to the sum
of Coulomb and nuclear contributions calculated separately. 
This can be seen in Fig.(\ref{fig:sum}) where the sum of the Coulomb and nuclear
cross sections for the CDCC calculation is compared
with the CDCC cross section when Coulomb and nuclear are treated in the same footing.
As in \cite{nunes1,shyam98} there is a wide range of angles where 
the Coulomb-nuclear interference effects cannot be neglected.

So far we have included all possible couplings within the subspace considered.
However it is useful to identify the relative importance of the 
continuum-continuum couplings as compared to the couplings to and from the ground state.
In Fig.(\ref{fig:coup}) we show the results of calculating the full
multi-step breakup into s and p bin-states, including continuum-continuum
couplings (dark lines) and excluding them (light lines). As can be seen, 
the continuum-continuum couplings are responsible for the
significant cross section reduction, which are not merely due
to depletion of flux from the elastic channel.  The reduction is still obtained
in calculations (not shown) with E2 couplings acting to only first order,
but not in those with E1 couplings only to first order. This indicates
that the reduction is caused by multistep E1 processes interfering
with low-order E2 transitions, a process similar to that seen in ref. \cite{esb}.

In order to elucidate the different $^8$B partial wave contributions,
we show in Fig.(\ref{fig:partial}) the differential cross section
obtained for the full multi-step breakup including: s (dotted), s+p
(short-dashed), s+p+d (long-dashed) and s+p+d+f (dot-dashed) bin
states. A good description of the physics can be obtained without d and
f-waves, although if one wishes to extract quantitative results these
should be included together with a finer energy-bin grid.

One of the main motivations of $^8$B breakup experiments is astrophysical,
to determine the $S_{17}$ at low relative energies. It is thus
important to disentangle the dependence on the $^8$B structure model.
This was the main concern of our earlier work \cite{puri}. In
Fig.(\ref{fig:kim}) we present a comparison of differential cross
sections for two $^8$B models: that of Kim \cite{kim}, and
our initial model, from Esbensen \cite{esb}.  The latter has a smaller
radius in order to reproduce $S_{17} \simeq 17$ eV b.  The difference between
the multi-step results using the two structure models is generally similar to
the difference in the 1-step results: there is an overall normalisation
due to the size of the projectile but no significant shape
change. For this particular case, the Kim model produces a $30 \%$
increase in the total differential cross section.

The Notre Dame group is considering repeating their experiment
\cite{nd} with a heavier  target at the same beam energy, hoping then
the $E2$ contribution will be easier to extract. 
We have therefore performed one and multistep calculations for
a $^{208}$Pb target, for all combinations of nuclear and Coulomb mechanisms
(see Fig.\ref{fig:pb1}). We have taken
the $^{208}$Pb - proton optical interaction from \cite{mako}. For the optical
potential between $^7$Be and $^{208}$Pb we have used a heavy-ion
global parameterisation  \cite{hions}. We have checked that the differential
cross section is not sensitive to variations on the core-target optical parameters, 
and thus we expect these results to give a good indication of the physical effects.  
In order to have quantitative results, measurements of the elastic scattering of $^7$Be 
(or a nucleus in the same mass/charge region) on $^{208}$Pb at these low
energies would be necessary.

Our results show that the nuclear contribution is zero up to $50^\circ$
and becomes important only at backward angles.
Given these results, it should be possible to extract information on
the magnitude of the electromagnetic components, as long as the detectors are
placed at smaller angles. One should keep in mind that  the
multi-step processes  reduce the Coulomb peak and alter its
shape.  

Similarly to what was found for $^{58}$Ni and 1-step DWBA, 
for a $^{208}$Pb target there are interference effects that do not allow 
a simple subtraction of the E2 component, as one would wish to obtain the $S_{17}$. 
In Fig.(\ref{fig:pb2}) we show the results for the CDCC calculations for the different
electric components together with the full calculation.  One can
clearly see a destructive interference between E1 and E2 components.
It may be possible to disentangle these components, but the result will
inevitably be model dependent.

\section{Conclusions}

Multistep calculations of low energy $^8$B breakup on $^{58}$Ni and $^{208}$Pb,
including all relevant couplings, have been performed for the first time.  We
have found it necessary to treat nuclear and Coulomb potentials on
the same footing, since there is considerable interference between
these mechanisms. Our calculations use the CDCC method, with Pad\'e approximants to
resum the Born series for the S-matrix.  Exact coupled-channel
calculations, possible in a reduced subspace, verify that
this Pad\'e resummation converges to the full coupled channel result.
We compare our results with the one-step prior-form DWBA cross sections
reported previously \cite{nunes1,shyam98}.

The multi-step effects are very strong, producing significant
reductions of the cross section compared with those from first order
theory. For Coulomb breakup we see pronounced interference effects for
all angles at and beyond the peak position, while multi-step effects are much
stronger for the nuclear part, so that, for the $^{58}$Ni target,
the nuclear peak resulting from
the 1-step calculations virtually disappears. The prior-form DWBA thus
overestimates the nuclear breakup probabilities at our
sub-Coulomb incident energy.  The multi-step reduction is principally 
due to the continuum-continuum couplings, not just to depletion
of the elastic channel, and this indicates
that the projectile undergoes considerable dynamical distortion and
recombination during the reaction.  The dominant qualitative changes
caused by the multi-step effects are seen when including only s and p
wave continuum bins, but d and f waves must be included 
for quantitative results.

The results for two different $^8$B structure models show
that the multi-step effects depend on the size of the projectile, and
are not sensitive to other details of the $^8$B g.s. wave function.
 With a $^{208}$Pb target, the nuclear contribution is only significant
for backward angles.  Multi-step effects reduce the cross section and
change the shape slightly.  According to our results, 
for the extraction of the E2 component
this experiment seems more promising than that with the lighter target,  
although care should be taken to account for the strong
destructive Coulomb-Coulomb interference.

\acknowledgements
We thank Jeff Tostevin for helpful discussions and comments. 
We would like to express our gratitude to Eduardo Lopes
for helping out with many computational difficulties.
UK support from the EPSRC grant GR/J/95867
and Portuguese support from JNICT PRAXIS/PCEX/P/FIS/4/96 are acknowledged.
One of the authors, F. Nunes, was supported by JNICT BIC 1481.

\newpage

\begin{figure}[htb]
\centerline{
	\parbox[t]{.7\textwidth}{
	\psfig{figure=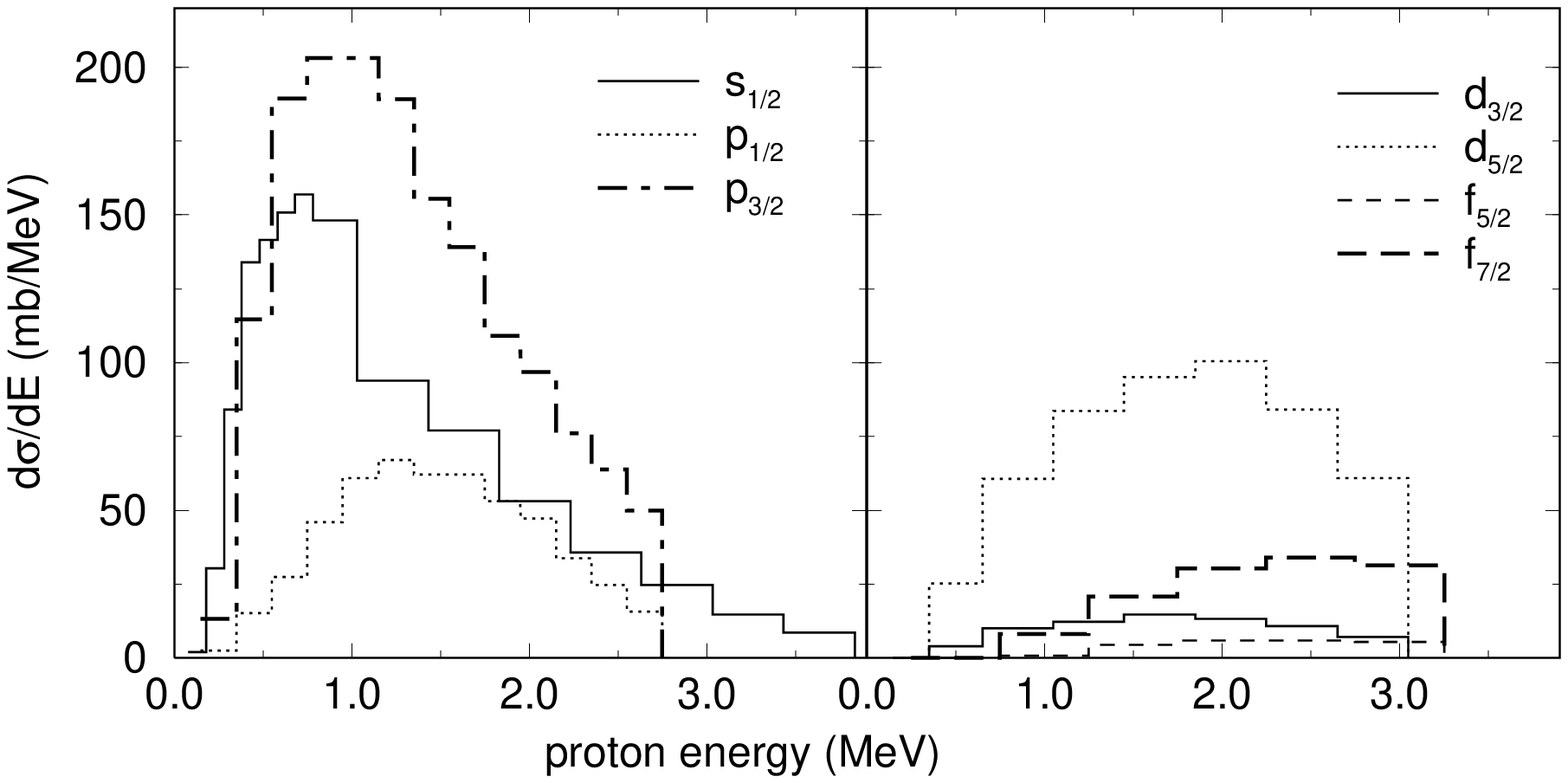,width=.7\textwidth}
	\caption{Energy distribution of s, p, d and f partial wave bins.}
	\label{fig:bin}}
}
\end{figure}
\begin{figure}[htb]
\centerline{
	\parbox[t]{0.7\textwidth}{
\centerline{\psfig{figure=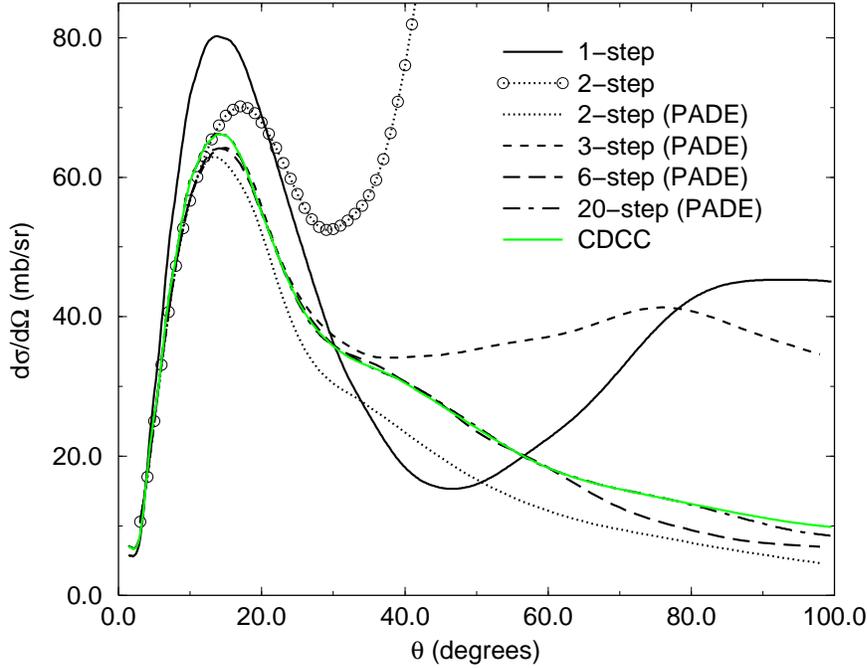,width=0.7\textwidth}}
	\caption{The differential cross section obtained for multi-step breakup
	of $^8$B into s and p-wave bins, including both Coulomb and nuclear
	effects: the full CDCC calculation,
the 1-step and 2-step DWBA, and higher order calculations using Pad\'e acceleration.}	
\label{fig:steps}}
}
\end{figure}
\begin{figure}[htb]
\centerline{
	\parbox[t]{0.7\textwidth}{
\centerline{\psfig{figure=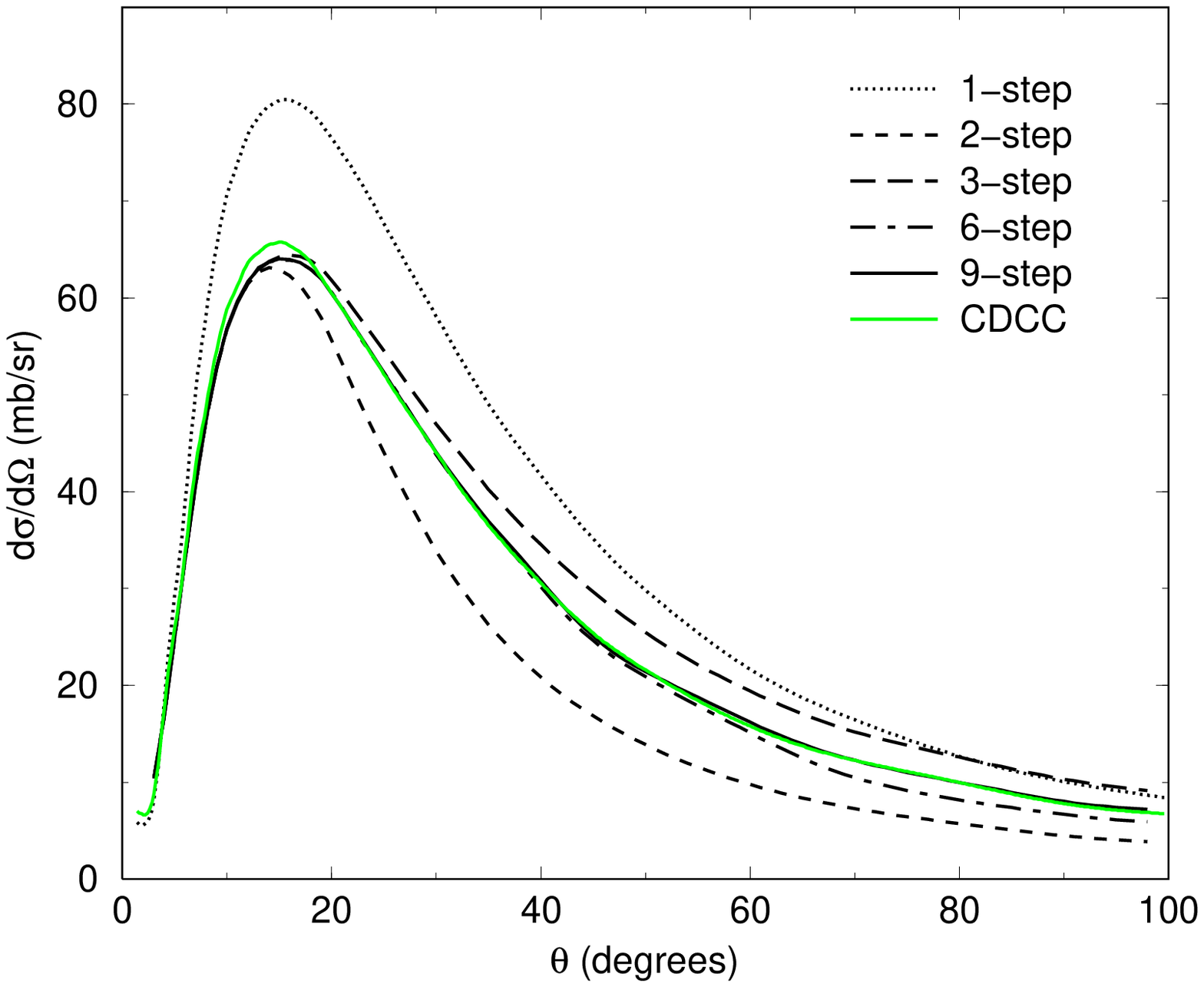,width=0.7\textwidth}}
	\caption{The differential cross section obtained for the
	multi-step Coulomb breakup of $^8$B into s and p-wave bins using Pad\'e acceleration.}	
\label{fig:stepc}}
}
\end{figure}
\begin{figure}[htb]
\centerline{
	\parbox[t]{0.7\textwidth}{
\centerline{\psfig{figure=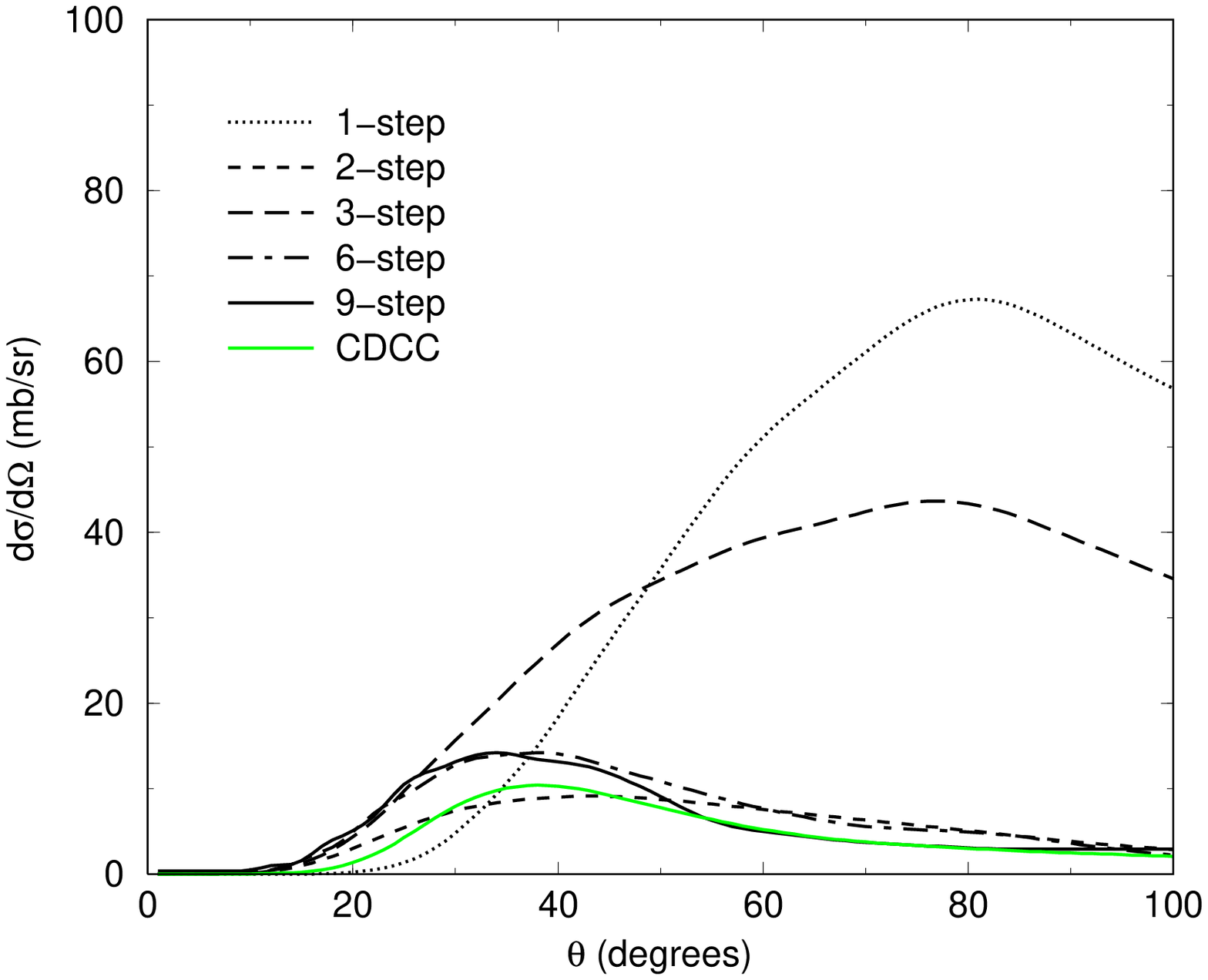,width=0.7\textwidth}}
	\caption{The differential cross section obtained for the
	multi-step nuclear breakup of $^8$B into s and p-wave bins, 
	using Pad\'e acceleration.}	
\label{fig:stepn}}
}
\end{figure}
\begin{figure}[htb]
\centerline{
	\parbox[t]{0.7\textwidth}{
\centerline{\psfig{figure=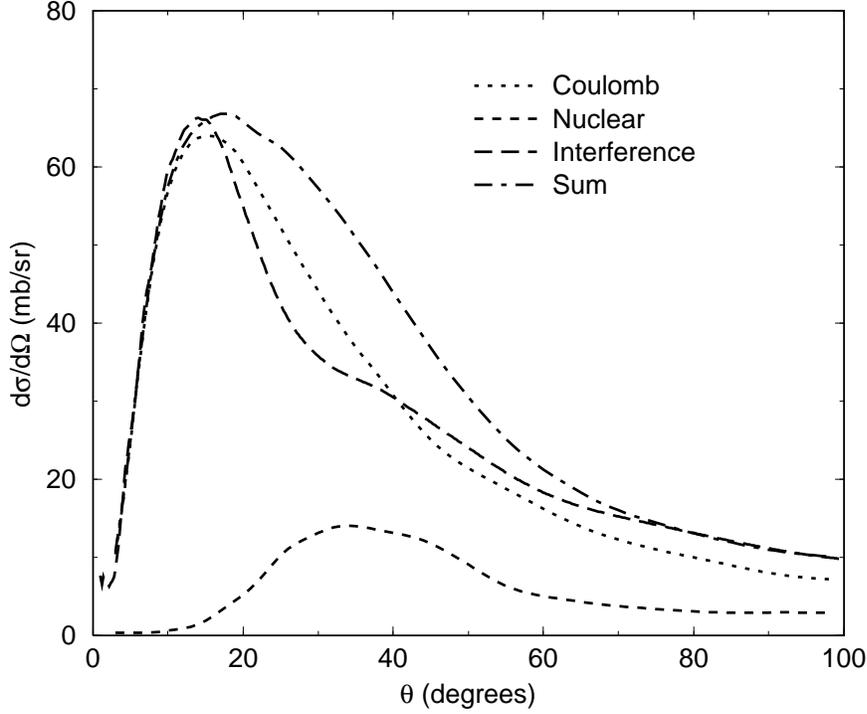,width=0.7\textwidth}}
	\caption{The CDCC differential cross section obtained for the breakup of
$^8$B into s-wave and p-wave bins:
comparison of the Coulomb and nuclear summed cross section with
the  calculation that includes nuclear-Coulomb interference. }	
\label{fig:sum}}
}
\end{figure}
\begin{figure}[htb]
\centerline{
	\parbox[t]{0.7\textwidth}{
\centerline{\psfig{figure=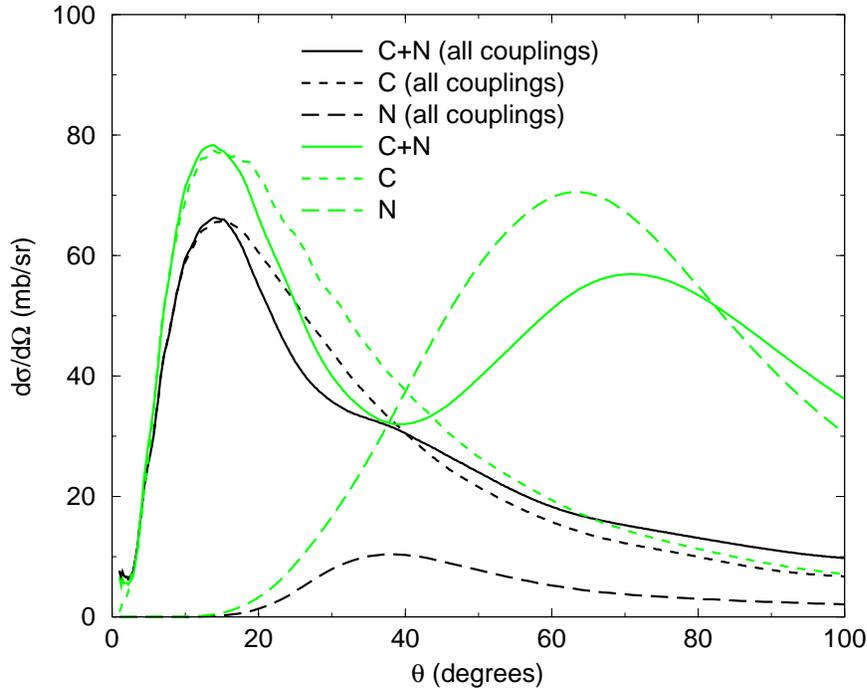,width=0.7\textwidth}}
	\caption{Comparing the CDCC differential cross section when 
no continuum-continuum couplings are included in the calculation
to the full calculation (this calculation includes s and p waves only).}	
\label{fig:coup}}
}
\end{figure}
\begin{figure}[htb]
\centerline{
	\parbox[t]{0.7\textwidth}{
\centerline{\psfig{figure=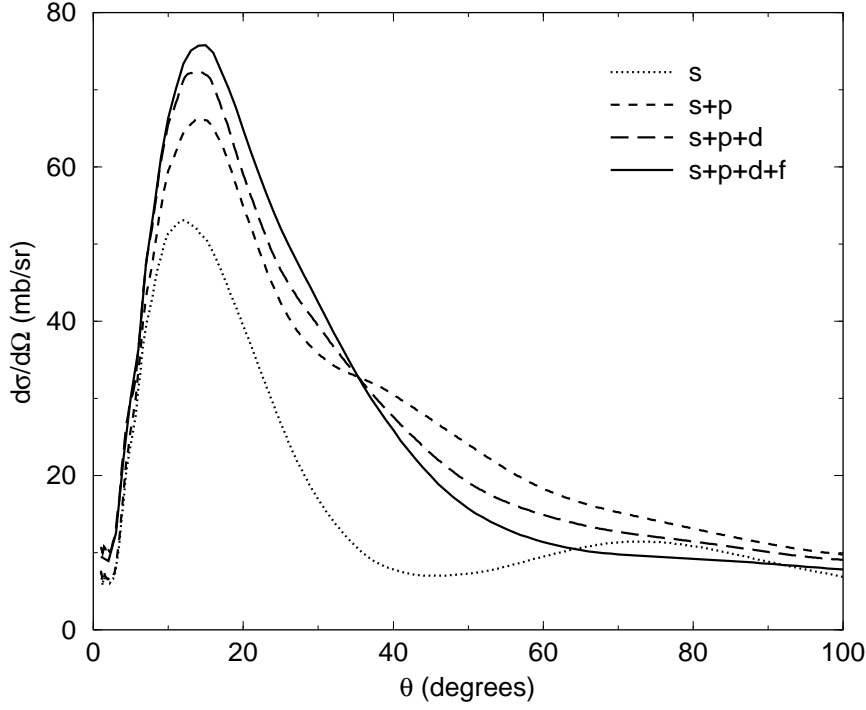,width=0.7\textwidth}}
	\caption{Cumulative $^8$B partial wave contributions to the
	CDCC differential cross section for Coulomb plus nuclear breakup to the 
	l=0,1,2 and 3 channels.}	
\label{fig:partial}}
}
\end{figure}
\begin{figure}[htb]
\centerline{
	\parbox[t]{0.7\textwidth}{
\centerline{\psfig{figure=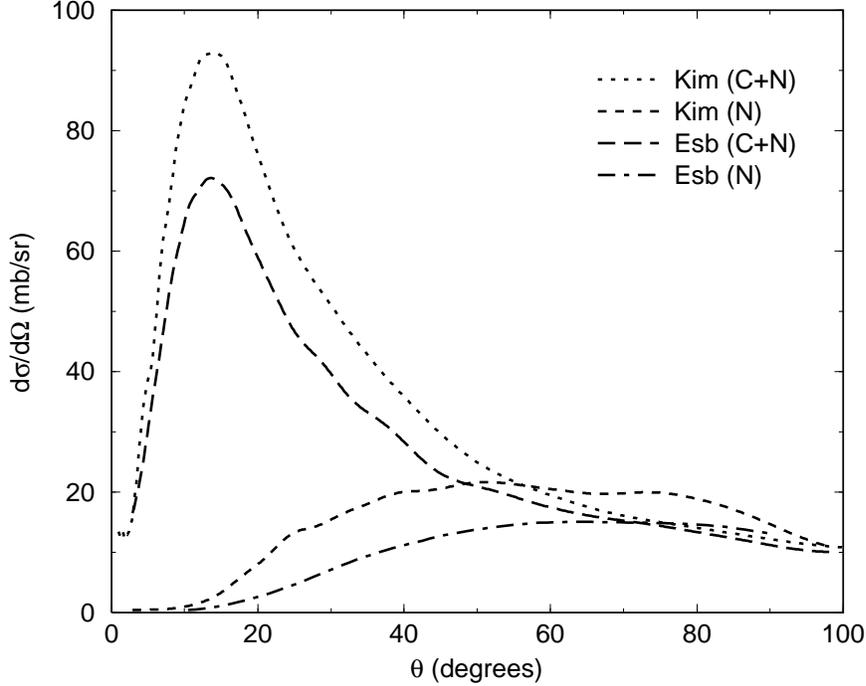,width=0.7\textwidth}}
	\caption{Sensitivity to the $^8$B structure model: the 
	CDCC differential cross section including all but f-wave bins.}	
\label{fig:kim}}
}
\end{figure}
\begin{figure}[htb]
\centerline{
	\parbox[t]{0.7\textwidth}{
\centerline{\psfig{figure=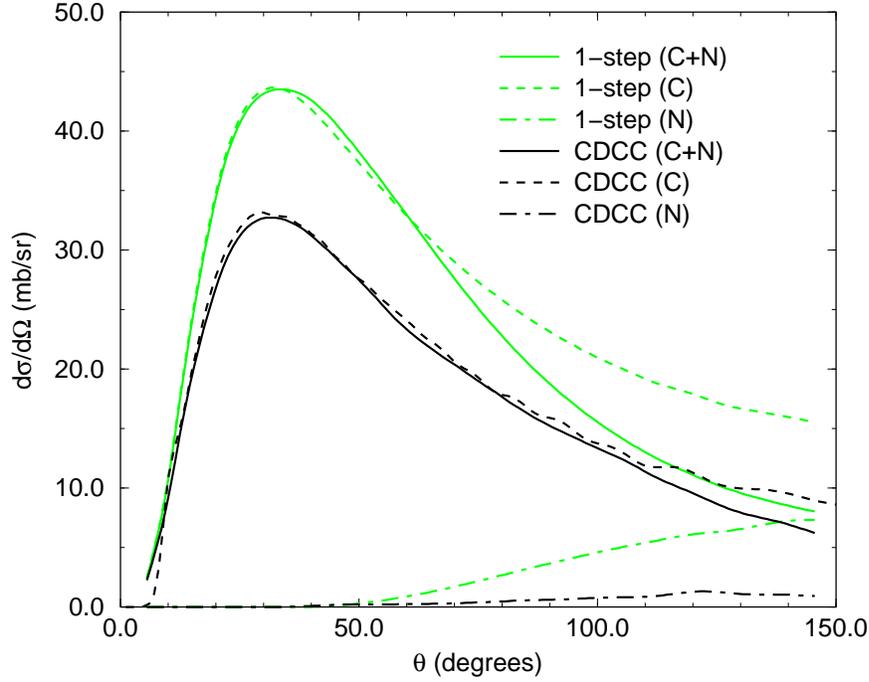,width=0.7\textwidth}}
	\caption{Comparing the 1-step and CDCC differential cross section for $^8$B 
	breakup on $^{208}$Pb, including all but f-wave bins.}	
\label{fig:pb1}}
}
\end{figure}
\begin{figure}[htb]
\centerline{
	\parbox[t]{0.7\textwidth}{
\centerline{\psfig{figure=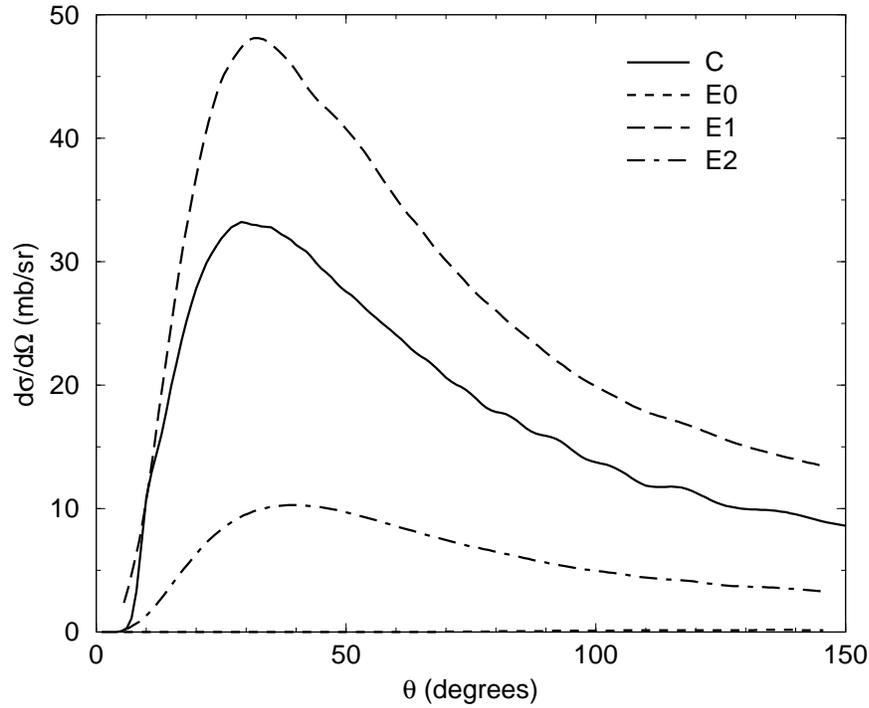,width=0.7\textwidth}}
	\caption{Comparing the CDCC effects of only E0, only E1,  and only E2 with 
the full  Coulomb differential cross section 
for $^8$B Coulomb breakup on $^{208}$Pb, including all but f-wave bins.}	
\label{fig:pb2}}
}
\end{figure}

\end{document}